\documentclass[runningheads]{llncs}
\usepackage{graphicx}
\usepackage{xcolor}
\usepackage{multirow}
\usepackage{amsfonts}
\usepackage{amsmath}
\usepackage[flushleft]{threeparttable}
\usepackage{textcomp}
\usepackage{ulem}

\usepackage[ruled,linesnumbered]{algorithm2e}

\begin{document}
%
\title{Modeling Dynamic Heterogeneous Network for Link Prediction using Hierarchical Attention with Temporal RNN}
\titlerunning{Dynamic Heterogeneous Network Embedding via DyHATR}
%
\author{Hansheng Xue\inst{1} \and Luwei Yang\inst{2}\thanks{Equal contribution.} \and Wen Jiang\inst{2} \and Yi Wei\inst{2}  \and Yi Hu\inst{2} \and Yu Lin\inst{1}\thanks{Corresponding author.}}
%
\authorrunning{Xue et al.}
%
\institute{Research School of Computer Science, The Australian National University, Canberra, Australia \\ \email{\{hansheng.xue, yu.lin\}@anu.edu.au} \and
Alibaba Group, Hangzhou, China \\
\email{\{luwei.ylw, wen.jiangw, yi.weiy, erwin.huy\}@alibaba-inc.com}
}
%
\maketitle              
\begin{abstract}
Network embedding aims to learn low-dimensional representations of nodes while capturing structure information of networks. It has achieved great success on many tasks of network analysis such as link prediction and node classification. 
Most of existing network embedding algorithms focus on how to learn static homogeneous networks effectively. 
However, networks in the real world are more complex, e.g., networks may consist of several types of nodes and edges (called heterogeneous information) and may vary over time in terms of dynamic nodes and edges (called evolutionary patterns). 
Limited work has been done for network embedding of dynamic heterogeneous networks as it is challenging to learn both evolutionary and heterogeneous information simultaneously. 
In this paper, we propose a novel dynamic heterogeneous network embedding method, termed as DyHATR, which uses hierarchical attention to learn heterogeneous information and incorporates recurrent neural networks with temporal attention to capture evolutionary patterns. 
We benchmark our method on four real-world datasets for the task of link prediction. Experimental results show that DyHATR significantly outperforms several state-of-the-art baselines.

\end{abstract}
\section{Introduction}
Network embedding is to encode network structures into non-linear space and represent nodes of networks as low-dimensional features~\cite{Cai2018Survey,Cui2019Survey}. It has been a popular and critical machine learning task with wide applications in many areas, such as recommender systems~\cite{Shi2019HERec,Wen2018WWW}, natural language processing~\cite{tu2017cane,Fang2016CQA} and computational biology~\cite{Nelson2019,Gligorijevic2017deepNF,peng2019bib}.

Existing network embedding methods have achieved significant performance on many downstream tasks such as link prediction and node classification, most of these approaches focus on static homogeneous networks~\cite{Perozzi2014DeepWalk,Grover2016node2vec,Tang2015LINE}, static heterogeneous networks~\cite{Dong2017metapath2vec,zhang2018MNE,Xue2019DeepMNE,Cen2019GATNE} or dynamic homogeneous networks~\cite{goyal2018dyngem,Goyal2019dyngraph2vec,zhou2018dynamicTriad,zhang2018timers,sankar2018DySAT}. However, many networks in the real-world are both dynamic and heterogeneous, which usually contain multiple types of nodes or edges~\cite{Dong2017metapath2vec,Cen2019GATNE} and the structure of networks may evolve over time~\cite{Du2018IJCAI,Trivedi2019DyRepLR}. For example, customer-product networks are usually heterogeneous with multiple node types to distinguish customers and products as well as dynamic with evolving nodes and edges to capture dynamic user activities.

Network embedding is challenging for dynamic heterogeneous networks to capture both heterogeneous information and evolutionary patterns. Dynamic networks are usually described as an ordered list of static network snapshots~\cite{Singer2019tNodeEmbed,Pareja2020EvolveGCN,Chen2019elstmd,Li2018DeepDN,Lu2019MMDNE}. As nodes and edges may vary in different snapshots, network embedding over dynamic heterogeneous networks need not only to capture the structural information of static heterogeneous snapshots but also to learn the evolutionary patterns between consecutive snapshots. 

Currently, limited work has been done for network embedding of dynamic heterogeneous networks. MetaDynaMix~\cite{Fard2019MetaDynaMix} integrates metapath-based topology features and latent representations to learn both heterogeneity and temporal evolution. 
Change2vec~\cite{Bian2019Change2Vec} focuses on measuring changes in snapshots instead of learning the whole structural information of each snapshot, and it also uses a metapath-based model to capture heterogeneous information. 
The above two methods both focus on short-term evolutionary information between adjacent snapshots of dynamic networks and thus become insufficient to capture long-term evolutionary patterns. 
More recently, Sajadmanesh et al.~\cite{sajadmaneshtkdd19} uses a recurrent neural network model to learn long-term evolutionary patterns of dynamic networks on top of metapath-based models and proposes a non-parametric generalized linear model, NP-GLM, to predict continuous-time relationships. 
Yin et al.~\cite{Yin2019DHNENR} propose a DHNE method, which learns both the historical and current heterogeneous information and models evolutionary patterns by constructing comprehensive historical-current networks based on consecutive snapshots. Then, DHNE performs metapath-based random walk and dynamic heterogeneous skip-gram model to capture representations of nodes. 
Kong et al.~\cite{Kong2019LinkPO} introduce a dynamic heterogeneous information network embedding method called HA-LSTM. It uses graph convolutional network to learn heterogeneous information networks and employs attention model and long-short time memory to capture evolving information over timesteps. 
Refer to Table \ref{tab1} for a brief summary of existing network embedding methods. 

To better capture the heterogeneity of static snapshots and model evolutionary patterns among consecutive snapshots. In this paper, we propose a novel dynamic heterogeneous network embedding method, named DyHATR. Specifically, DyHATR uses a hierarchical attention model to learn static heterogeneous snapshots and temporal attentive RNN model to capture evolutionary patterns. The contributions of this paper are summarized as:

\begin{itemize}
    \item We propose a novel dynamic heterogeneous network embedding method, named DyHATR, which captures both heterogeneous information and evolutionary patterns.
    \item We use the hierarchical attention model (including node-level and edge-level attention) to capture the heterogeneity of static snapshots.
    \item We emphasize the evolving information of dynamic networks and use the temporal attentive GRU/LSTM to model the evolutionary patterns among continuous snapshots.
    \item We evaluate our method DyHATR on the task of link prediction. The results show that DyHATR significantly outperforms several state-of-the-art baselines on four real-world datasets.
\end{itemize}

\begin{table}[ht!]
\caption{Summary of typical network embedding methods.}\label{tab1}
\resizebox{\textwidth}{34mm}{
\begin{tabular}{|c|c|c|c|c|}
\hline
\bf{Embedding Type} & \bf{Method} & \bf{Node Type} & \bf{Edge Type} & \bf{Dynamic} \\
\hline
 & DeepWalk~\cite{Perozzi2014DeepWalk} & single & single & No \\
 & node2vec~\cite{Grover2016node2vec} & single & single & No \\
Static Homogeneous Network & LINE~\cite{Tang2015LINE} & single & single & No \\
Embedding (SHONE) & GCN~\cite{kipf2017semi} & single & single & No \\
 & GraphSAGE~\cite{hamilton2017inductive} & single & single & No \\
 & GAT~\cite{velickovic2018graph} & single & single & No \\
\hline
 \multirow{2}{*}{Static Heterogeneous Network} & metapath2vec~\cite{Dong2017metapath2vec} & multiple & single & No \\
 \multirow{2}{*}{Embedding (SHENE)}& MNE~\cite{zhang2018MNE} & single & multiple & No \\
 & GATNE-T~\cite{Cen2019GATNE} & multiple & multiple & No \\
\hline
 & DynGEM~\cite{goyal2018dyngem} & single & single & Yes \\
 & DynamicTriad~\cite{zhou2018dynamicTriad} & single & single & Yes \\
Dynamic Homogeneous Network & dyngraph2vec~\cite{Goyal2019dyngraph2vec} & single & single & Yes \\
Embedding (DHONE) & EvolveGCN~\cite{Pareja2020EvolveGCN} & single & single & Yes \\
 & E-LSTM-D~\cite{Chen2019elstmd} & single & single & Yes \\
 & DySAT~\cite{sankar2018DySAT} & single & single & Yes \\
\hline
\multirow{4}{*}{Dynamic Heterogeneous Network} & MetaDynaMix~\cite{Fard2019MetaDynaMix} & multiple & single & Yes \\
\multirow{4}{*}{Embedding (DHENE)} & HA-LSTM~\cite{Kong2019LinkPO} & multiple & single & Yes \\
 & change2vec~\cite{Bian2019Change2Vec} & multiple & multiple & Yes \\
 & NP-GLM~\cite{sajadmaneshtkdd19} & multiple & multiple & Yes \\
 & DHNE~\cite{Yin2019DHNENR} & multiple & multiple & Yes \\
\hline
\end{tabular}}
\end{table}

\begin{figure}
\centerline{\includegraphics[width=0.975\textwidth]{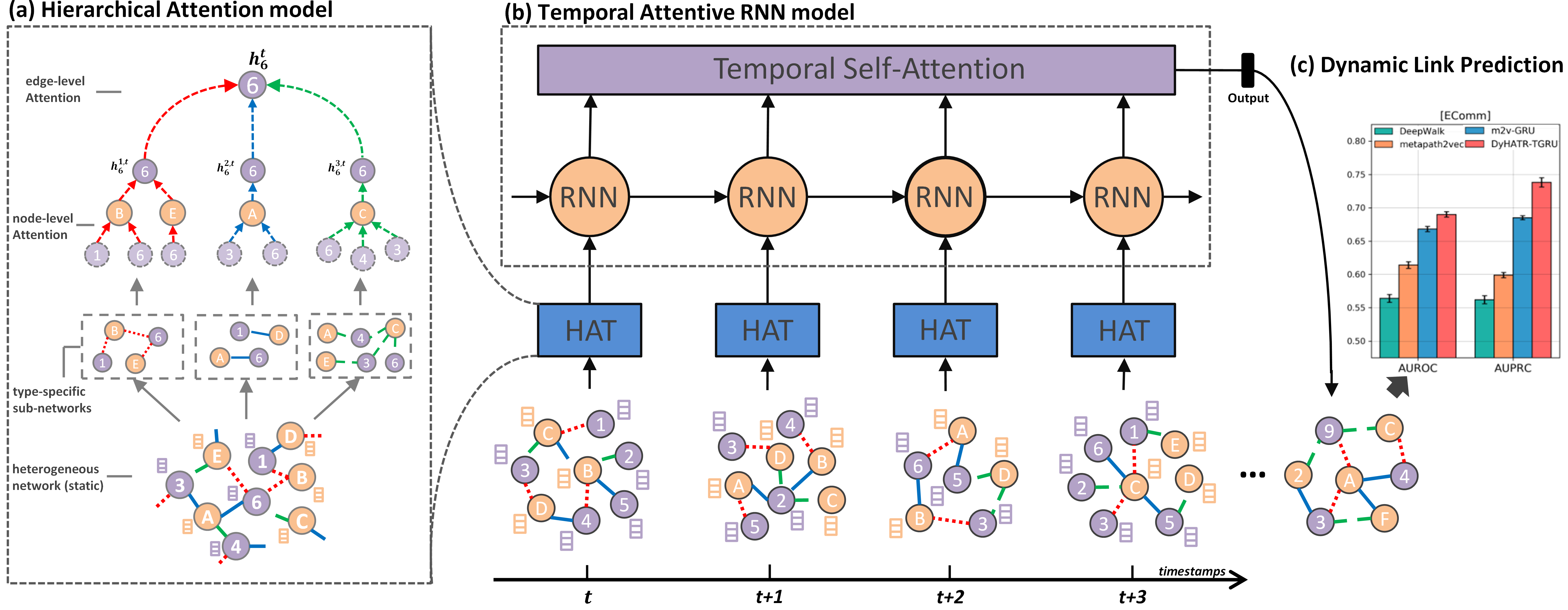}}
\caption{The overall framework of DyHATR. The proposed model contains three parts, (a) Hierarchical Attention model, (b) Temporal Attentive RNN model and (c) Dynamic Link Prediction. The right histogram shows the performance improvement
of DyHATR over DeepWalk, metapath2vec, and metapath2vec-GRU on the EComm dataset.}\label{framework}
\end{figure}

\section{Method}
The main task of dynamic heterogeneous network embedding is how to capture heterogeneous information and temporal evolutionary patterns over dynamic networks simultaneously. We propose a novel dynamic heterogeneous network embedding method that uses the hierarchical attention model to capture the heterogeneity of snapshots and temporal attentive RNN model to learn the evolutionary patterns over evolving time. The whole framework of our proposed model, DyHATR, is shown in Figure~\ref{framework}.

Dynamic heterogeneous networks are often described as an ordered list of observed heterogeneous snapshots $\mathbb{G}= \{G^1,G^2,...,G^T\}$, where $G^t=(V^t,E^t)$ represents the $t$-th snapshot. 
$V^t$ is the node set with node type $o \in O$, $E^t$ is the edge set with edge type $r \in R$. $O$ and $R$ are node type and edge type set respectively, and $|O| + |R| > 2$. $T$ is the number of snapshots. 
The aim of dynamic network embedding is to learn a non-linear mapping function that encodes node $v\in V^t$ into a latent feature space $f: v \to \mathbf{z}^t_v$, where $\mathbf{z}^t_v \in \mathbb{R}^d, d\ll |V^t|$. For each snapshot $G^t$, the learned embedding matrix can be represented as $\mathbf{Z}^t\in \mathbb{R}^{|V^t|\times d}$. 
Note that each snapshot $G^t$ is considered as a static heterogeneous network and we will first explain how DyHATR uses the hierarchical attention model to capture the heterogeneity of snapshots in the next section. Then the temporal attentive RNN for modeling evolutionary patterns across snapshots follows.

\subsection{Hierarchical Attention for Heterogeneous Information}
We introduce hierarchical attention model to capture heterogeneity information of static snapshots by splitting heterogeneous snapshots into several type-specific sub-networks according different edge types. The hierarchical attention model contains two components, node-level attention and edge-level attention. The illustration of hierarchical attention is shown in Figure~\ref{framework} (a).

\subsubsection{Node-level Attention} The node-level attention model aims to learn the importance weight of each node's neighborhoods and generate novel latent representations by aggregating features of these significant neighbors. 
%
%
For each static heterogeneous snapshot $G^t\in \mathbb{G}$, we employ attention models for every subgraph with the same edge type. The weight coefficient of node pair $(i,j)$ for edge type $r$ and $t$-th snapshot can be calculated as:
\begin{equation}\label{eq:node_level_attention_1}
\alpha^{rt}_{i,j}=\frac{\exp(\sigma(\mathbf{a}_r^\top \cdot [\mathbf{W}^{r}\cdot \mathbf{x}_i||\mathbf{W}^{r}\cdot \mathbf{x}_j]))}{\sum_{k\in N^{rt}_i}\exp(\sigma(\mathbf{a}_r^\top \cdot [\mathbf{W}^{r}\cdot \mathbf{x}_i||\mathbf{W}^{r}\cdot \mathbf{x}_k]))}
\end{equation}
where $\mathbf{x}_i$ is the initial feature vector of node $i$, $\mathbf{W}^{r}$ is a transformation matrix for edge type $r$. $N^{rt}_i$ is the sampled neighbors of node $i$ with edge type $r$ in the snapshot $t$; $\mathbf{a}_r$ is the parameterized weight vector of attention function in $r$-th edge type; $\sigma$ is the activation function; and $||$ denotes the operation of concatenation. Then, aggregating the neighbors' latent embedding with calculated weight coefficients, we can obtain the final representation of node $i$ at edge type $r$ and $t$-th snapshot as
\begin{equation}\label{eq:node_level_attention_2}
\hat{\mathbf{h}}^{rt}_i=\sigma(\sum_{j\in N_i^{rt}}\alpha^{rt}_{ij}\cdot \mathbf{W}^{r} \mathbf{x}_j)
\end{equation}
where $\hat{\mathbf{h}}^{rt}_i$ is the aggregated embedding of node $i$ for edge type $r$ and $t$-th snapshot. To obtain stable and effective features, we employ the node-level attention model with multi-head mechanisms. Specifically, we parallelly run $\kappa$ independent node-level attention models and concatenate these learned features as the output embedding. Thus, the multi-head attention representation of node $i$ at edge type $r$ and $t$-th snapshot can be described as
\begin{equation}\label{eq:node_level_attention_3}
    \mathbf{h}^{rt}_i=\textnormal{Concat}(\hat{\mathbf{h}}^1,\hat{\mathbf{h}}^2,...,\hat{\mathbf{h}}^\kappa)
\end{equation}
where $\hat{\mathbf{h}}^{\kappa}$ is the simplified symbol of $\hat{\mathbf{h}}^{rt}_i$, which is the output from the $\kappa$-th node-level attention model, and $\kappa$ is the number of attention heads. After the multi-head node-level attention model, we can obtain node representation sets with different edge type and snapshot, $\{\mathbf{h}_1^{rt}, \mathbf{h}_2^{rt},...,\mathbf{h}_{|V^t|}^{rt}, \mathbf{h}_i^{rt}\in\mathbb{R}^E\}$, where $E\ll |V^t|$ is the dimension of the embedded features at the node-level attention model. 

\subsubsection{Edge-level Attention}
Node-level attention model can capture single edge-type-specific information. However, heterogeneous networks usually contain multiple types of edges. To integrate multiple edge-specific information for each node, we employ an edge-level attention model to learn the importance weight of different types of edges and aggregate these various type-specific information to generate novel embeddings. 

Firstly, we input edge-specific embeddings into non-linear transformation functions to map into the same feature space $\sigma (\mathbf{W}\cdot \mathbf{h}^{rt}_i+ \mathbf{b})$, where $\sigma$ is the activation function (i.e., tanh function); $\mathbf{W}$ and $\mathbf{b}$ are learnable weight matrix and bias vector respectively. Both these parameters are shared across all different edge types and snapshots. Then, we measure the importance coefficients of inputting edge-specific embedding by calculating similarities between mapped edge-specific embedding and edge-level attention parameterized vector $\mathbf{q}$. The normalized weight coefficients of node $i$ for edge type $r$ and $t$-th snapshot can be represented as $\beta^{rt}_i$. Then, we can aggregate these edge-specific embeddings to generate the final representation features of node $i$ at $t$-th snapshot:

\begin{equation}\label{eq:edge_level_attention_1}
    \beta_i^{rt} = \frac{\exp(\mathbf{q}^\top \cdot \sigma(\mathbf{W} \cdot \mathbf{h}_i^{rt}+\mathbf{b}))}{\sum_{r \in R} \exp(\mathbf{q}^\top \cdot\sigma(\mathbf{W} \cdot \mathbf{h}_i^{rt}+\mathbf{b}))} \qquad \mathbf{h}_i^t = \sum_{r=1}^R\beta_i^{rt}\cdot \mathbf{h}_i^{rt}
\end{equation}  



After aggregating all edge-specific embeddings, we can obtain the final node embeddings for different snapshots, $\{\mathbf{h}_1^t, \mathbf{h}_2^t,...,\mathbf{h}_{|V^t|}^t, \mathbf{h}_i^t\in\mathbb{R}^F \}$, where $F\ll |V^t|$ is the dimension of the embeddings outputted by edge-level attention model.

\subsection{Temporal Attentive RNN for Evolutionary Patterns}
The temporal evolutionary patterns show the appearance and disappearance of nodes and edges as time goes by. The hierarchical attention model of DyHATR can capture the heterogeneity of static snapshots effectively, but it cannot model patterns evolving over time. Recently, the introduction of the recurrent neural network (RNN) for the dynamic network embedding methods achieve promising performance~\cite{Goyal2019dyngraph2vec,Singer2019tNodeEmbed,Xu2019STAR,sankar2018DySAT}. 
In our paper, we extend the existing RNN model and propose a temporal attentive RNN model to capture deeper evolutionary patterns among continuous timestamps. The proposed temporal attentive RNN model mainly contains two parts, recurrent neural network and temporal self-attention model. The illustration of temporal attentive RNN model is shown in Figure~\ref{framework} (b). 

\subsubsection{Recurrent Neural Network Model} 
The structure of recurrent neural network enables the capability of modeling
sequential information and learning evolutionary patterns among continuous snapshots. Two prominent and widely-used variants of RNN are used in our proposed model, Long short-term memory (LSTM) and Gated-recurrent-unit (GRU). 

Long short-term memory (LSTM) is an classical recurrent neural network model which can capture long-distance information of a sequence data and achieves promising performance on dynamic network learning tasks~\cite{Hochreiter1997LongSM}. Through the above hierarchical attention model, we can obtain node embedding set for each snapshot $\{\mathbf{h}_1^t, \mathbf{h}_2^t,...,\mathbf{h}_{|V^t|}^t, \mathbf{h}_i^t\in\mathbb{R}^F$ \}, where $t$ is the $t$-th snapshot, $i$ indicates the node $i$, and $F$ is the dimension of each node embedding. 
Each LSTM unit computes the input vector $\mathbf{i}^t$, the forget gate vector $\mathbf{f}^t$, the output gate vector $\mathbf{o}^t$, the memory cell vector $\mathbf{c}^t$ and the state vector $\mathbf{s}^t$. The formulations of single LSTM unit are as follows:

\begin{equation}\label{eq:lstm}
\begin{aligned} 
  &  \mathbf{i}^t = \sigma(\mathbf{W}_i\cdot[\mathbf{h}^t || \mathbf{s}^{t-1}] + \mathbf{b}_i), \\
  &  \mathbf{f}^t = \sigma(\mathbf{W}_f\cdot[\mathbf{h}^t || \mathbf{s}^{t-1}] + \mathbf{b}_f), \\
  &  \mathbf{o}^t = \sigma(\mathbf{W}_o\cdot[\mathbf{h}^t || \mathbf{s}^{t-1}] + \mathbf{b}_o), \\
  &  \widetilde{\mathbf{c}^t} = \tanh(\mathbf{W}_c\cdot[\mathbf{h}^t ||  \mathbf{s}^{t-1}] + \mathbf{b}_c), \\
  &  \mathbf{c}^t = \mathbf{f}^t \odot \mathbf{c}^{t-1} + \mathbf{i}^t \odot \widetilde{\mathbf{c}^t}, \\
  &  \mathbf{s}^t = \mathbf{o}^t\odot \tanh(\mathbf{c}^t),
\end{aligned}
\end{equation}

where $t\in\{1,2,...,T\}$; $\mathbf{i}^t,\mathbf{f}^t,\mathbf{o}^t,\mathbf{c}^t \in\mathbb{R}^{D}$ are input, forget, output and memory cell gates respectively; $\mathbf{W}_i,\mathbf{W}_f,\mathbf{W}_o,\mathbf{W}_c\in \mathbb{R}^{D\times2F}$ and $\mathbf{b}_i,\mathbf{b}_f,\mathbf{b}_o,\mathbf{b}_c\in\mathbb{R}^D$ are trainable parameters; $\sigma$ is the activation function; $||$ denotes the operation of concatenation; and $\odot$ is an element-wise multiplication operator. 

Gated-recurrent-unit (GRU) is a simpler variant of LSTM~\cite{Cho2014LearningPR} and introduced to capture the long-term association in many dynamic network embedding models~\cite{Xu2019STAR,Pareja2020EvolveGCN}. 
GRU units have fewer parameters to learn compared to LSTM units, and each GRU unit only contains two gate cells, update gate and reset gate. The state vector $\mathbf{s}^t\in \mathbb{R}^D$ for each input snapshot can be calculated as the following equations iteratively:

\begin{equation}\label{eq:gru}
\begin{aligned} 
  &  \mathbf{z}^t = \sigma(\mathbf{W}_z\cdot[\mathbf{h}^t || \mathbf{s}^{t-1}] + \mathbf{b}_z), \\
  &  \mathbf{r}^t = \sigma(\mathbf{W}_r\cdot[\mathbf{h}^t || \mathbf{s}^{t-1}] + \mathbf{b}_r), \\
  &  \widetilde{\mathbf{s}^t} = \tanh(\mathbf{W}_s\cdot[\mathbf{h}^t || (\mathbf{r}^t \odot \mathbf{s}^{t-1})] + \mathbf{b}_s), \\
  &  \mathbf{s}^t = (1-\mathbf{z}^t)\odot \mathbf{s}^{t-1} + \mathbf{z}^t \odot\widetilde{\mathbf{s}^t},
\end{aligned}
\end{equation}
where $t\in\{1,2,...,T\}$; $\mathbf{z}^t,\mathbf{r}^t \in\mathbb{R}^{D}$ are update and reset gates respectively; $\mathbf{W}_z,\mathbf{W}_r,\mathbf{W}_s\in \mathbb{R}^{D\times2F}$ and $\mathbf{b}_z,\mathbf{b}_r,\mathbf{b}_s\in\mathbb{R}^D$ are trainable parameters; $\sigma$ is the activation function; $||$ denotes the operation of concatenation; and $\odot$ is an element-wise multiplication operator. In the iterative equation of GRU, we use $\mathbf{r}^t\odot \mathbf{s}^{t-1}$ to model the latent features communicating from the previous state. 

The outputs of the RNN model are denoted as $\{\mathbf{s}^t_1,\mathbf{s}^t_2,...,\mathbf{s}^t_{|V^t|}, \mathbf{s}^t_i\in\mathbb{R}^D$ \}. Then, the traditional step may concatenate these state vectors $[\mathbf{s}^1_i || \mathbf{s}^2_i ||,..., || \mathbf{s}^T_i]$ or take the last state $\mathbf{s}^T_i$ as the final embedding for node $i$. However, these traditional ways may lead to information loss and cannot capture the most important embedding features. In our model, we employ a temporal-level attention model on the output of the RNN model to capture the significant feature vectors.

\subsubsection{Temporal Self-Attention model} 
Instead of concatenating all feature vectors together as the final embedding to predict dynamic links in the last snapshot, we employ a temporal-level self-attention model to further capture the evolutionary patterns over dynamic networks. From the output of RNN model, we can obtain the embeddings of node $i$ at different snapshots, $\{\mathbf{s}_i^1,\mathbf{s}_i^2,...,\mathbf{s}_i^T, \mathbf{s}_i^t\in\mathbb{R}^D\}$, where $D$ is the dimension of input embeddings. Assuming the output of temporal-level attention model is $\{\mathbf{z}_i^1,\mathbf{z}_i^2,...,\mathbf{z}_i^T\}, \mathbf{z}_i^t\in\mathbb{R}^{D^{\prime}}$, where $D^{\prime}$ is the dimension of output embeddings.

The Scaled Dot-Product Attention is used in our model to learn the embeddings of each node at different snapshot. We pack representations from previous step for node $i$ across time, $\mathbf{S}_i\in\mathbb{R}^{T\times D}$, as input. The corresponding output is denoted as $\mathbf{Z}_i\in\mathbb{R}^{T\times D^{\prime}}$. Firstly, the input is mapped into different feature space, $\mathbf{Q}_i=\mathbf{S}_i\mathbf{W}_q$ for queries, $\mathbf{K}_i=\mathbf{S}_i\mathbf{W}_k$ for keys, and
$\mathbf{V}_i=\mathbf{S}_i\mathbf{W}_v$ for values, where $\mathbf{W}_q,\mathbf{W}_k,\mathbf{W}_v\in{\mathbb{R}^{D\times D^{\prime}}}$ are trainable parameters. The temporal-level attention model is defined as:

\begin{equation}\label{eq:temporal_level_attention_1}
    \mathbf{Z}_i=\mathbf{\Gamma}_i \cdot
    \mathbf{V}_i=\textnormal{softmax}(\frac{(\mathbf{S}_i\mathbf{W}_q)(\mathbf{S}_i\mathbf{W}_k)^T}{\sqrt{D^{'}}}+\mathbf{M})\cdot (\mathbf{S}_i\mathbf{W}_v)
\end{equation}

where $\mathbf{\Gamma}_i\in\mathbb{R}^{T\times T}$ is the importance matrix; $\mathbf{M}\in\mathbb{R}^{T\times T}$ denotes the mask matrix. If $M_{uv}=-\infty$, it means the attention from snapshot $u$ to $v$ is off and the corresponding element in the importance matrix is zero, i.e., $\Gamma_i^{uv}=0$. When snapshot satisfies $u\leq v$, we set $M_{u,v}=0$; otherwise $M_{u,v}=-\infty$. 

It is also feasible to use multi-head attention to improve the effectiveness and robustness of the model. The final emebddings of node $i$ 
with $\kappa^{\prime}$ heads can be formulated as $\mathbf{Z}_i=\textnormal{Concat}(\mathbf{\hat{Z}}_i^1,\mathbf{\hat{Z}}_i^2,...,\mathbf{\hat{Z}}_i^{\kappa^{\prime}})$. At this step, we obtain all of the embeddings of node $i$ across all snapshots. But we will use embeddings of the last snapshot, denoted as $\mathbf{z}_i^T$, for optimization and downstream task.

\subsection{Objective Function}
The aim of DyHATR is to capture both heterogeneous information and evolutionary patterns among dynamic heterogeneous networks, we define the objective function that utilizes a binary cross-entropy function to ensure node $u$ at the last snapshot have similar embedding features with its nearby nodes. The loss function of DyHATR is defined as:
\begin{equation}\label{eq:cost_function}
\small
\begin{gathered}
\textit{L}(\textbf{z}_u^T) = \sum_{v\in N^T(u)}-\log(\sigma(<\mathbf{z}_u^T, \mathbf{z}_v^T>))-Q\cdot\mathbb{E}_{v_n\sim\textit{P}_n(v)}\log(\sigma(<-\mathbf{z}_u^T,\mathbf{z}_{v_n}^T>)) + \lambda\cdot\textit{L}_{p}
\end{gathered}
\end{equation}
where $\sigma$ is the activation function (i.e., sigmoid function); $<.>$ is the inner product operation; $N^T(u)$ is the neighbor with fixed-length random walk of node $u$ at the last $t$-th snapshot; $\textit{P}_n(v)$ denotes the negative sampling distribution and $Q$ is the number of negative samples; $\textit{L}_{p}$ is the penalty term of the loss function to avoid over-fitting, i.e., $L_2$ regularization; $\lambda$ is the hyper-parameters to control the penalty function. 
The pseudocode for DyHATR is shown in Algorithm 1. 

\begin{algorithm}
\caption{The DyHATR algorithm.}
\label{alg:dyhatr}
\KwIn{A sequence of snapshots $\mathbb{G}= \{G^1,G^2,...,G^T\}$ with $G^t=(V^t,E^t)$, the number of embedding layers $d$, the number of multi-heads $\kappa$, the number of neighbor samples $k$, initialization parameters\;}
\KwOut{Node Embedding $\mathbf{Z}$}
\For{each timestamp $t \in [1,2,...T]$}{
Construct edge-type-specific sub-networks set $R$\;
\For{each edge-type $r \in R$}{
Obtain the weight coefficient of node pair $(i,j)$, $\alpha^{rt}_{i,j}$ by Equation (1)\;
Obtain the representation of node $i$ at edge type $r$ and $t$-th snapshot, $\hat{\mathbf{h}^{rt}_i}$ by Equation (2)\;
The multi-head attention of node $i$, $\mathbf{h}^{rt}_i\leftarrow \textnormal{Concat}(\hat{\mathbf{h}}^1,\hat{\mathbf{h}}^2,...,\hat{\mathbf{h}}^\kappa)$\;
}
Calculate the weight coefficient of node $i$ for edge type $r$ and $t$-th snapshot, $\beta_i^{rt}$\; 
Obtain the representation of node $i$ at $t$-th snapshot, $\mathbf{h}_i^t$ by Equation (4)\;
}
Obtain $\mathbf{s}^t_i$ by Equation (5) or (6) through GRU/LSTM \;
Get $\mathbf{Z}_i$ by Equation (7) through temporal self-attention model\;
Optimize the loss function $\textit{L}(\textbf{z}_u^T)$ by Equation (8)\; 

\textbf{return} $\mathbf{Z}$;
\end{algorithm}

\section{Experiments}
\subsection{Datasets}
We benchmark DyHATR on dynamic link prediction to evaluate the performance of network embedding. Four real-world dynamic network datasets are used in our experiments. The statistics of these datasets are listed in Table~\ref{tab2}. 

\textbf{Twitter.}\footnote{http://snap.stanford.edu/data/higgs-twitter.html} This social network is sampled from the Higgs Twitter dataset of the SNAP platform\cite{snapnets}. It reflects the three kinds of user behaviors (re-tweeting, replying and mentioning) between 1st and 7th July 2012.

\textbf{Math-Overflow.}\footnote{http://snap.stanford.edu/data/sx-mathoverflow.html} This temporal network is collected from the stack exchange website Math Overflow and opened on the SNAP platform~\cite{snapnets}. It represents three different types of interactions between users (answers to questions,
comments to questions, comments to answer) within 2,350 days. In our experiments, we split this time span into 11 snapshots.

\textbf{EComm.}\footnote{https://tianchi.aliyun.com/competition/entrance/231719} This dataset, real-world heterogeneous bipartite graphs of the e-commerce, is extracted from the AnalytiCup challenge of CIKM-2019. EComm mainly records shopping behaviors of users within 11 daily snapshots from 10th June 2019 to 20th June 2019, and it consists of two types of nodes (users and items) and four types of edges (click, buy, add-to-cart and add-to-favorite).

\textbf{Alibaba.com.} This dataset consists of user behavior logs collected from the e-commerce platform of Alibaba.com\cite{yang2019DyHAN}. It mainly contains customers' activities records from 11th July 2019 to 21st July 2019, and consists of two types of nodes (users and items) and three types of activities (click, enquiry and contact).

\begin{table}[htbp]
\centering
\caption{Statistics of four real-world datasets.}\label{tab2}
\begin{tabular}{|c|c|c|c|c|c|}
\hline
 Dataset & \ \#Nodes \ & \ \#Edges \ & \#Node Types & \#Edge Types & \#snapshots \\
\hline
 Twitter & 100,000 & 63,410 & 1 & 3 & 7 \\
 Math-Overflow & 24,818 & 506,550 & 1 & 3 & 11 \\
 EComm & 37,724 & 91,033 & 2 & 4 & 11 \\
 Alibaba.com \ & 16,620 & 93,956 & 2 & 3 & 11 \\
\hline
\end{tabular}
\end{table}

\subsection{Experimental Setup}
The task of dynamic link prediction is to learn node representations over previous $t$ snapshots and predict the links at $(t+1)$-th snapshot. 
Specifically, the previous $t$ snapshots $\{G^1,...,G^t\}$ are used to learn representations of nodes and the $(t+1)$-th snapshot $G^{t+1}$ is the whole evaluation set. We randomly sample 20\% of edges in the evaluation set (snapshot $G^{t+1}$) as the hold-out validation set to tune hyper-parameters. Then the remaining 80\% of edges in the snapshot $G^{t+1}$ are used for link prediction task. Among the remaining evaluation set, we further randomly select $25\%$ edges and the rest $75\%$ edges as training and test sets respectively for link prediction. Meanwhile, we randomly sample equal number of pairs of nodes without link as negative examples for training and test sets respectively. We use the inner product of the embedding features of two nodes as the feature of the link.

For this task, we train a Logistic Regression model as the classifier, which is similar with previous works~\cite{zhou2018dynamicTriad,sankar2018DySAT}. We use the area under the ROC curve (AUROC) and area under the precision-recall curve (AUPRC) as evaluation metrics. We repeatedly run our model and baselines for five times and report the mean and standard deviation values. 

The proposed DyHATR model is conducted on four real-world datasets using the Linux server with 6 Intel(R) Core(TM) i7-7800X CPU @3.50GHz, 96GB RAM and 2 NVIDIA TITAN Xp 12GB. The codes of DyHATR are implemented in Tensorflow 1.14 and Python 3.6. 
In the DyHATR model, we use multi-head attention model with 4 or 8 heads to capture node feature embeddings. The dimension of the final embedding is 32. The stochastic gradient descent and Adam optimizer are used in our proposed model to update and optimize parameters. In the link prediction part, we use a logistic regression classifier and evaluation metric functions from the scikit-learn library. 
%
DyHATR is freely available at https://github.com/skx300/DyHATR.

\subsection{Baselines}
We compare DyHATR with several state-of-the-art network embedding methods, including three for static homogeneous networks (DeepWalk~\cite{Perozzi2014DeepWalk}, GraphSAGE~\cite{hamilton2017inductive} and GAT~\cite{velickovic2018graph}), one for static heterogeneous networks (metapath2vec~\cite{Dong2017metapath2vec}), and three for dynamic homogeneous networks (DynamicTriad~\cite{zhou2018dynamicTriad}, dyngraph2vec~\cite{Goyal2019dyngraph2vec} and DySAT~\cite{sankar2018DySAT}). For static network embedding methods, we firstly integrate all snapshots into a static network. Then, we apply them on the integrated static network to learn latent representations of nodes. 

There exists four dynamic heterogeneous network embedding methods, MetaDynaMix~\cite{Fard2019MetaDynaMix}, change2vec~\cite{Bian2019Change2Vec}, DHNE~\cite{Yin2019DHNENR} and NP-GLM~\cite{sajadmaneshtkdd19}. Here, we mainly focus on DHNE and NP-GLM method. The representative algorithm NP-GLM mainly uses the idea of combining metapath2vec and RNN to learn dynamic heterogeneous networks, we thus compare DyHATR with two methods, metapath2vec-GRU and metapath2vec-LSTM. To make a fair comparison, we set the same final embedding size as DyHATR, i.e. 32, for all baselines. The hyper-parameters for different baselines are all optimized specifically to be optimal. 

\begin{table}
\caption{The experimental results on the task of dynamic link prediction. The bold value is the best performance achieved by DyHATR. The underline indicates the highest AUROC/AUPRC score achieved by baselines.}\label{results_all}
\begin{threeparttable}
\resizebox{\textwidth}{28mm}{
\begin{tabular}{|c|c|c|c|c|c|c|c|c|}
\hline
 \multirow{2}{*}{\bf{\scriptsize{Methods}}} & \multicolumn{2}{c|}{\bf{\scriptsize{Twitter}}} & \multicolumn{2}{c|}{\bf{\scriptsize{Math-Overflow}}} & \multicolumn{2}{c|}{\bf{\scriptsize{EComm}}} & \multicolumn{2}{c|}{\bf{\scriptsize{Alibaba.com}}}\\
 \cline{2-9}
  & \scriptsize{AUROC} & \scriptsize{AUPRC} & \scriptsize{AUROC} & \scriptsize{AUPRC} & \scriptsize{AUROC} & \scriptsize{AUPRC} & \scriptsize{AUROC} & \scriptsize{AUPRC} \\
\hline
 \scriptsize{DeepWalk} & 0.520(0.040) & 0.686(0.021) & 0.714(0.002) & 0.761(0.001) & 0.564(0.006) & 0.562(0.006) & 0.546(0.019) & 0.570(0.008) \\
 \scriptsize{GraphSAGE-mean} & 0.600(0.024) & 0.765(0.016) & 0.683(0.005) & 0.712(0.003)) & 0.635(0.007) & 0.642(0.008) & 0.571(0.019) & \uline{0.610(0.014)} \\
 \scriptsize{GraphSAGE-meanpool} & 0.562(0.009) & 0.711(0.005) & 0.671(0.007) & 0.697(0.008) & 0.600(0.018) & 0.590(0.017) & 0.552(0.015) & 0.570(0.021) \\
 \scriptsize{GraphSAGE-maxpool} & 0.571(0.031) & 0.731(0.021) & 0.637(0.008) & 0.660(0.009) & 0.594(0.010) & 0.587(0.009) & 0.542(0.009) & 0.573(0.008) \\
 \scriptsize{GraphSAGE-LSTM} & 0.546(0.016) & 0.710(0.010) & 0.672(0.005) & 0.697(0.008) & 0.599(0.013) & 0.588(0.015) & 0.567(0.006) & 0.585(0.007) \\
 \scriptsize{GAT} & 0.587(0.019) & 0.744(0.013) & \uline{0.737(0.004)} & 0.768(0.003) & 0.648(0.003) & 0.642(0.008) & \uline{0.572(0.010)} & \uline{0.610(0.008)} \\
\hline
 \scriptsize{metapath2vec} & 0.520(0.040) & 0.686(0.021) & 0.714(0.002) & 0.761(0.001) & 0.614(0.005) & 0.599(0.004) &  0.561(0.004) & 0.601(0.007) \\
\hline
 \scriptsize{DynamicTriad} & 0.626(0.027) & 0.785(0.013) & 0.704(0.005) & 0.759(0.003) & 0.614(0.008) & \uline{0.688(0.007)} & 0.543(0.013) & 0.598(0.011) \\ 
 \scriptsize{dyngraph2vec-AE} & 0.549(0.025) & 0.724(0.021) & 0.517(0.006) & 0.565(0.005) & 0.503(0.002) & 0.509(0.006) & 0.506(0.019) & 0.534(0.014) \\
 \scriptsize{dyngraph2vec-AERNN} & 0.511(0.034) & 0.706(0.031) & 0.582(0.003) & 0.598(0.003) & 0.503(0.004) & 0.501(0.004) & 0.505(0.008) & 0.554(0.045) \\
 \scriptsize{DySAT} & \uline{0.634(0.003)} & \uline{0.796(0.005)} & 0.506(0.008) & 0.543(0.006) & 0.512(0.002) & 0.513(0.002) & 0.521(0.013) & 0.550(0.010)\\
\hline
 \scriptsize{metapath2vec-GRU} & 0.539(0.025) & 0.744(0.018) & 0.730(0.004) & 0.770(0.004) & \uline{0.668(0.004)} & 0.685(0.003)  & 0.544(0.003) & 0.576(0.004) \\
 \scriptsize{metapath2vec-LSTM} & 0.554(0.027) & 0.754(0.017) & 0.727(0.004) & \uline{0.771(0.003)} & 0.666(0.005) & 0.683(0.008)  & 0.547(0.003) & 0.578(0.008) \\
 \scriptsize{DHNE} &0.552(0.005) &0.649(0.010) &0.678(0.001) &0.720(0.001) &0.547(0.003) &0.626(0.004) &0.515(0.001) &0.561(0.003) \\
\hline
 \scriptsize{DyHATR-TGRU} & 0.649(0.005) & 0.805(0.007) & 0.741(0.002) & 0.781(0.001) & 0.690(0.004) & \textbf{0.738(0.007)}\tnote{*}  &0.590(0.008)  &0.602(0.008)  \\
 \scriptsize{DyHATR-TLSTM} & \textbf{0.660(0.010)}\tnote{*} & \textbf{0.810(0.009)}\tnote{*} & \textbf{0.751(0.001)}\tnote{*} & \textbf{0.790(0.002)}\tnote{*} & \textbf{0.696(0.005)}\tnote{*} & 0.734(0.005)  &\textbf{0.605(0.003)}\tnote{*}  &\textbf{0.617(0.005)}  \\
\hline
\end{tabular}}
\begin{tablenotes}
\tiny{\item[*] Asterisks indicate the improvement over baselines achieved by DyHATR is significant (rank-sum p-value $<$ 0.01).}
\end{tablenotes}
\end{threeparttable}
\end{table}

\subsection{Experimental Results}
\textbf{Task of link prediction.} The experimental results are summarized in Table~\ref{results_all}. Overall, DyHATR achieves the best performance on both AUROC and AUPRC metrics over four datasets among all baselines, including two typical dynamic heterogeneous network embedding methods (metapath2vec-GRU and metapath2vec-LSTM). The highest AUROC and AUPRC achieved by DyHATR on the EComm dataset are 0.696 and 0.738 respectively, which are significantly higher than the scores of baselines (0.668 for AUROC by metapath2vec-GRU and 0.688 for AUPRC by DynamicTrid). For Twitter, the AUROC and AUPRC score of DyHATR-TLSTM are 0.660 and 0.810 respectively, which are sightly higher than the second highest score achieved by DySAT (0.634 for AUROC and 0.796 for AUPRC). Besides, we also perform the rank-sum test to validate the significance of experimental results achieved by DyHATR.
Note that dynamic homogeneous network embedding methods (such as DySAT and DynamicTriad) perform relatively better on the Twitter and Math-Overflow datasets than EComm and Alibaba.com datasets, probably because there are less number of node and edge types in Twitter and Math-Overflow than the other two datasets. 


\textbf{Effectiveness of Hierarchical Attention model.} The dynamic heterogeneous network can be represented as an ordered series of static heterogeneous snapshots. In our proposed DyHATR model, we use a hierarchical attention model to capture the heterogeneity of static snapshots. To evaluate the effectiveness of DyHATR on capturing heterogeneity of static snapshots, we compare the hierarchical attention model (HAT) with metapath2vec (m2v), a classic and widely-used heterogeneous network embedding method. 

\begin{figure}[tbp]
\centerline{\includegraphics[width=0.95\textwidth]{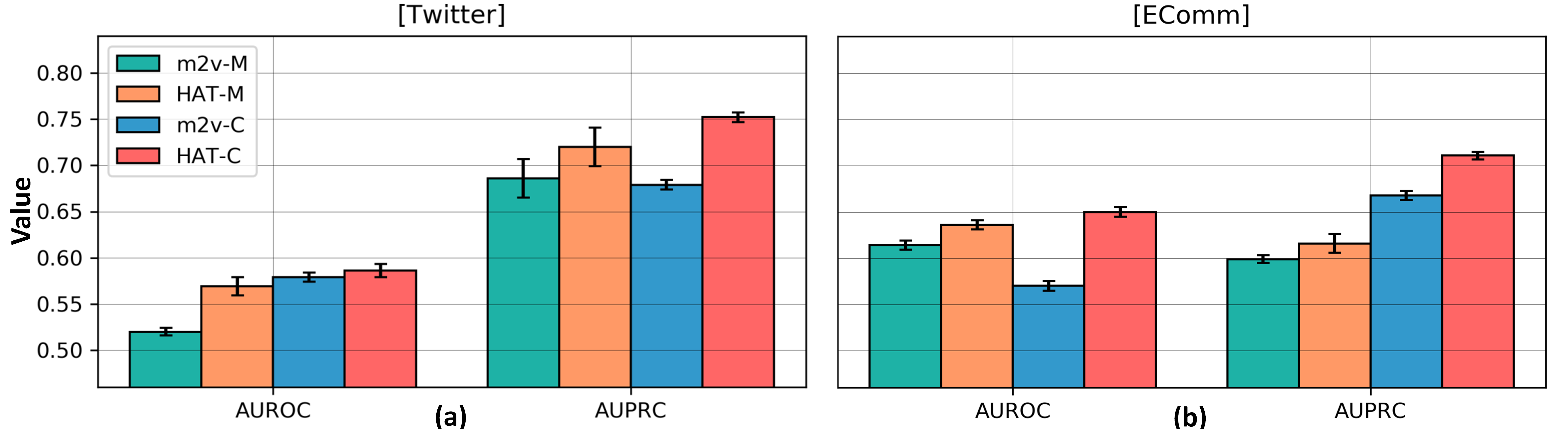}}
\caption{The comparison results between hierarchical attention model (HAT) and metapath2vec (m2v). The x-axis shows two evaluation metrics (AUROC and AUPRC). The y-axis represents the evaluation value. HAT-M/m2v-M denotes merge snapshots into a single network and run HAT/m2v model; HAT-C/m2v-C represents run HAT/m2v first on each snapshot and concatenate these embeddings to predict dynamic links.} \label{hat}
\end{figure}

To the full comparison, we adopt two different ways to integrate multiple snapshots. One is to merge snapshots into a single heterogeneous snapshot and run HAT/m2v representation learning methods. Another is to run HAT/m2v on each snapshot respectively and concatenate these embedding vectors for the downstream tasks. From the experimental results (Figure~\ref{hat}), the hierarchical attention model outperforms metapath2vec on both Twitter and EComm datasets with two different integration methods, demonstrating the effectiveness of HAT on learning heterogeneous information of each snapshot. 

\begin{figure}[tb]
\centerline{\includegraphics[width=0.95\textwidth]{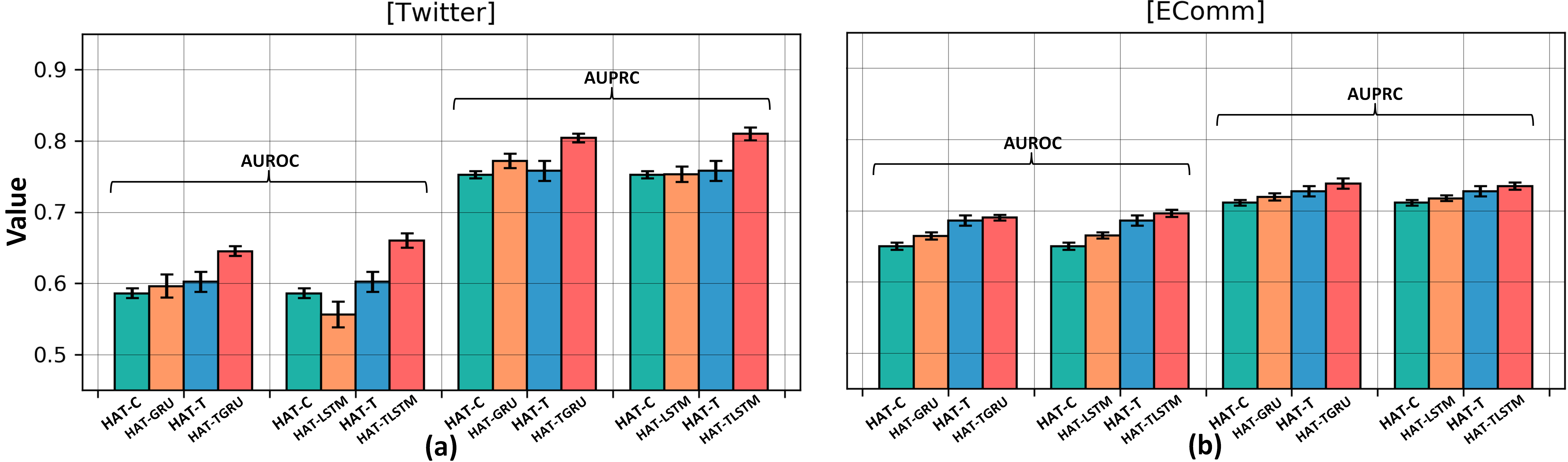}}
\caption{The comparison results of different components in DyHATR. The x-axis shows models with different components. The y-axis represents the AUROC/AUPRC value. HAT-C denotes hierarchical attention model; HAT-T is the HAT model with temporal attention model; TGRU/TLSTM denotes temporal attentive GRU/LSTM model.} \label{trnn}
\end{figure}

\textbf{Effectiveness of Temporal Attentive RNN model.} In our proposed method, DyHATR uses the temporal attentive GRU (TGRU) and LSTM (TSTM) model to learn the evolutionary patterns among continuous timestamps. 
%
To validate the effectiveness of this component. We have done an experiment by replacing/adding sub-component step by step. All methods use HAT to model each snapshot. The initial method is HAT-C, which concatenates the embedding vectors from each snapshot. Then the second method HAT-GRU/LSTM uses GRU/LSTM module to replace the concatenation to capture the sequential information. While the third one HAT-T uses temporal attention model instead. The final one HAT-TGRU/TLSTM, which is our proposed DyHATR, combines temporal attention model and GRU/LSTM. 
Figure~\ref{trnn} shows the experimental results on Twitter and EComm datasets. 
We can see that either GRU/LSTM or temporal attention model outperforms the naive concatenation. The combination of temporal attention model and GRU/LSTM further improve the performance better. This shows the superiority of temporal attentive RNN model on modeling evolving information.



\begin{figure}
\centerline{\includegraphics[width=0.975\textwidth]{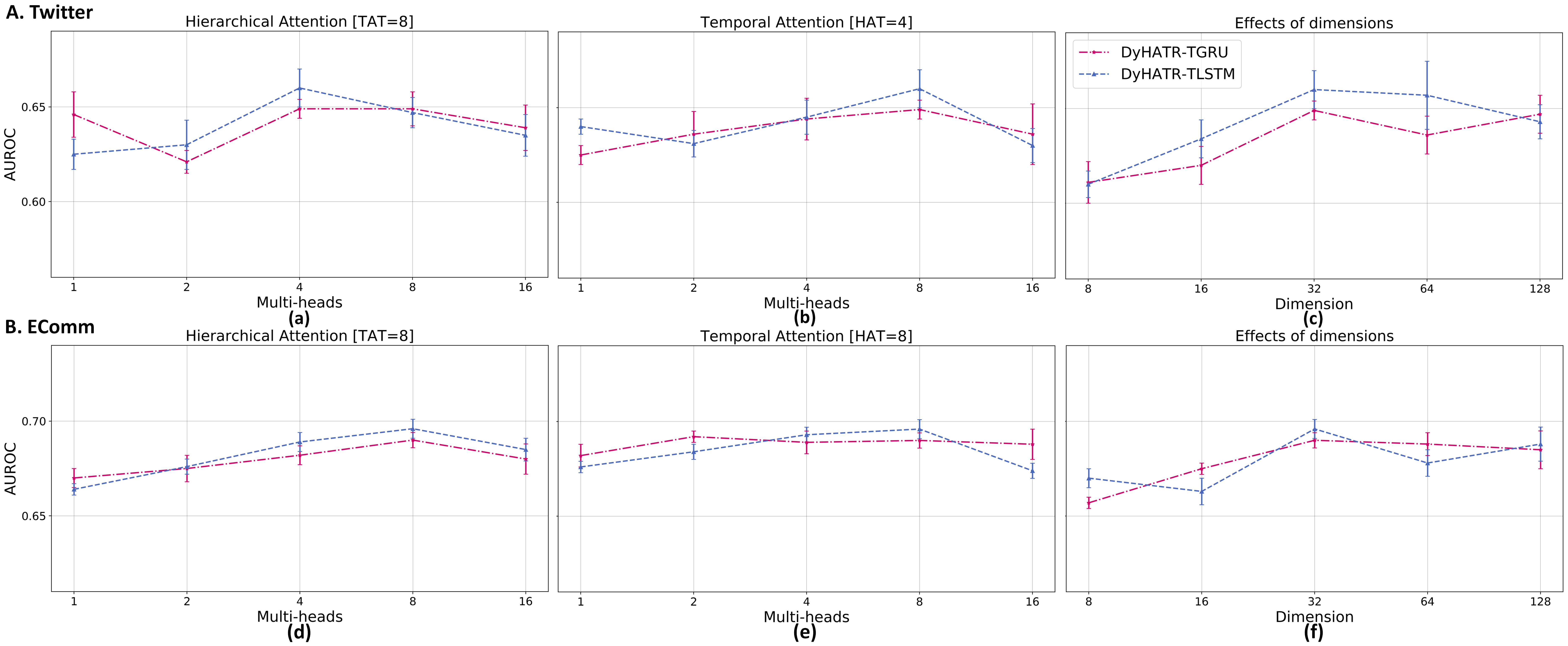}}
\caption{The results of DyHATR varying different parameters on Twitter and EComm dataset. Upper row shows the results on Twitter dataset. Lower row reports the results on EComm dataset. (a) and (d) varies the number of multi-heads on hierarchical attention; (b) and (e) varies the number of multi-heads on temporal attention model; (c) and (f) validates the effects of embedding dimensions. The x-axis shows the varying parameters and the y-axis represents the AUROC value.} \label{params}
\end{figure}

\textbf{Parameters Sensitivity.} 
The number of multi-heads in bot hierarchical attention (HAT) and temporal attention (TAT), and the dimension of final embedding output are critical parameters for our proposed DyHATR algorithm. Thus, we further analyze the effect of these parameters on the dynamic link prediction task. 
When we vary one parameter to check the sensitivity, the other parameters are kept fixed. Figure~\ref{params} (a) and (b) show the result on Twitter, the optimal numbers of multi-head for HAT and TAT are 4 and 8 respectively. Similarly, Figure~\ref{params} (d) and (e) report 8 and 8 for EComm.
%
Figure~\ref{params} (c) and (f) shows the performance respect to the dimension of final embedding output. When the dimension equals to 32, DyHATR achieves the highest AUROC value in our experiments.


Besides, we also compare the training time of DyHATR with metapath2vec-GRU/LSTM methods. For EComm, the training time of metapath2vec-GRU and metapath2vec-LSTM are 475s and 563s respectively. DyHATR is slightly slower than meatpath2vec--based methods (579s for DyHATR-TGRU  and 595s for DyHATR-TLSTM), because of the large scale of parameters in attention and RNN model. 

\textbf{Granularity of Snapshots.} In this paper, we describe dynamic network as an ordered list of snapshots. The duration of each snapshot will affect the total number of snapshots. For example, the duration of snapshot in EComm is one day, which results in 10 snapshots. We find that the finer granularity of snapshots, which means the shorter of the duration, would improve the performance. When the duration is set to two days in EComm, there are 5 snapshots in total. The AUROC values are 0.640 and 0.625 for DyHATR-TGRU and DyHATR-TLSTM respectively, which are worse than that of 10 snapshots. Same situation is observed for Twitter, when we set duration as two days, there are 3 snapshots. The AUROC values are 0.580 and 0.570 for DyHATR-TGRU and DyHATR-TLSTM respectively, which are worse than that of 6 snapshots. One of the reasons is that the finer granularity is more beneficial to capture evolving patterns.


\textbf{Effects of attributes.} Considering not all datasets contains node attributes, in previous experiments we only use node id embedding as the node input feature for convenience. However, it is easy to incorporate node inherent attributes as input feature.
For example, in EComm dataset, users and items contain many attributes, such as gender, age, education, career, income, category, brand, prize and so on. One may simply process and concatenate these information as input features of nodes.
Table~\ref{tab4} shows an experiment on EComm dataset using more attributes. The results show the scalability of our propsoed DyHATR and more attributes may improve performance.


\begin{table}[htbp]
\centering
\caption{The results of DyHATR without and with attributes on EComm dataset.}\label{tab4}
\begin{tabular}{|l|c|c|c|c|}
\hline
  \multirow{2}{*}{Node Input Feature} & \multicolumn{2}{|c|}{DyHATR-TGRU} & \multicolumn{2}{|c|}{DyHATR-TLSTM} \\
 \cline{2-5}
 & \scriptsize{AUROC} & \scriptsize{AUPRC} & \scriptsize{AUROC} & \scriptsize{AUPRC} \\
\hline
 node id & 0.690(0.004) & 0.738(0.007) & 0.696(0.005) & 0.734(0.005) \\
 node id + other attributes & 0.704(0.002) & 0.753(0.006) & 0.701(0.004) & 0.749(0.005) \\
\hline
\end{tabular}
\end{table}




\section{Conclusions}
Network embedding has recently been widely used in many domains and achieved significant progress. However, existing network embedding methods mainly focus on static homogeneous networks, which can not directly be applied to dynamic heterogeneous networks. In this paper, we propose a dynamic heterogeneous network embedding algorithm, termed as DyHATR, which can simultaneously learn heterogeneous information and evolutionary patterns. DyHATR mainly contains two components, hierarchical attention for capturing heterogeneity and temporal attentive RNN model for learning evolving information overtime. Compared with other state-of-the-art baselines, DyHATR demonstrates promising performance on the task of dynamic links prediction. Besides, our proposed DyHATR is scalable to attributed dynamic heterogeneous networks and model inherent information of nodes.

%
%
\bibliographystyle{splncs04}
\bibliography{sample}

\end{document}